\documentclass[conference]{IEEEtran}

\IEEEoverridecommandlockouts

\usepackage{cite}
\usepackage{overpic}
\usepackage{amsmath,amssymb,amsfonts}

\usepackage{graphicx}
\usepackage{textcomp}
\usepackage{xcolor}
\usepackage{float}
\usepackage{amsthm}
\usepackage{graphicx}
\usepackage{epstopdf}
\usepackage{amsmath,bm}
\usepackage{amsfonts}
\usepackage{amssymb}
\usepackage{color}
\usepackage{multirow}
\usepackage{multicol}
\usepackage{soul,xcolor}
\usepackage{algorithm}
\usepackage{algpseudocode}%

\usepackage{comment}
\usepackage{booktabs}
\usepackage{multirow}  
\usepackage{subcaption}

\theoremstyle{plain}

\newcommand{\vect}[1]{\mathbf{#1}}

\def\Htran{\mbox{\tiny $\mathrm{H}$}}
\def\Ttran{\mbox{\tiny $\mathrm{T}$}}
\def\CN{\mathcal{N}_{\mathbb{C}}} 
\def\imagunit{\mathsf{j}} 

\begin{document}

\title{Near-Field Channel Estimation with ELAA Modular Arrays Under Hardware Impairments \vspace{-0.2cm}
\thanks{The work by \"O. T. Demir was supported by 2232-B International Fellowship for Early Stage Researchers Programme funded by the Scientific and Technological Research Council of T\"urkiye. The work by E. Bj{\"o}rnson was supported by the  Grant 2022-04222 from the Swedish Research Council.}}

\author{\IEEEauthorblockN{ \"Ozlem Tu\u{g}fe Demir$^*$ and Emil Bj{\"o}rnson$^{\dagger}$}
\IEEEauthorblockA{{$^*$Department of Electrical-Electronics Engineering, TOBB University of Economics and Technology, Ankara, T\"urkiye
		} \\ {$^\dagger$Department of Computer Science, KTH Royal Institute of Technology, Stockholm, Sweden  
		} \\
		{Email: ozlemtugfedemir@etu.edu.tr, emilbjo@kth.se  \vspace{-0.4cm}}
}

}

\maketitle

\begin{abstract}
Extremely large-scale antenna arrays (ELAAs) enable high spatial resolution and multiplexing, especially for user equipments (UEs) in the radiative near-field. To reduce hardware cost, modular ELAA architectures with distributed baseband units (BBUs) are gaining traction. This paper addresses near-field line-of-sight (LOS) channel estimation under low noise amplifier (LNA)-induced hardware impairments in such modular systems. We propose computationally efficient estimators that exploit the array geometry and constant-modulus structure of near-field LOS channels, including a novel two-dimensional (2D) discrete Fourier transform (DFT) masking technique that improves estimation accuracy and significantly reduces fronthaul signaling. Numerical results show that the proposed methods significantly outperform the conventional least squares (LS) method.
\end{abstract}


        \vspace{-2mm}

\section{Introduction}
        \vspace{-2mm}

Extremely large-scale antenna arrays (ELAAs), composed of hundreds to thousands of antennas, have emerged as a key enabler for enhancing beamforming, spatial multiplexing, and spatial resolution in next-generation wireless systems \cite{hu2018beyond}. In particular, the large physical apertures of ELAAs create radiative near-field conditions, under which spherical wavefronts from different user equipments (UEs) become more distinguishable, thereby enhancing spatial multiplexing capabilities \cite{bjornson2024towards}. However, coordinating such large numbers of antennas incurs significant hardware complexity and, when baseband units (BBUs) are centralized, fronthaul signaling overhead. Instead of densely packing antennas into a monolithic array, modular array architectures—composed of widely spaced subarrays connected to distributed BBUs—offer a cost-effective and scalable alternative. By leveraging spatial sparsity and enabling local processing, these architectures reduce hardware complexity and interconnect demands while maintaining high performance \cite{kosasih2025near,li2023modular}.

In ELAAs, the cost of antennas becomes a critical concern due to the significantly larger number of elements compared to conventional systems. To keep deployment costs reasonable, low-cost hardware will likely be necessary, making hardware impairments inevitable. One notable impairment in the uplink arises from non-linearities introduced by the low-noise amplifiers (LNAs) in the RF chains of each antenna element~\cite{Jacobsson2018}. In this paper, we present the first study of near-field line-of-sight (LOS) channel estimation in ELAAs under LNA-induced nonlinearities. Estimating the near-field LOS channel requires searching over the UE’s location parameters, such as range, azimuth, and elevation. While a parametric maximum likelihood estimator could, in theory, exploit the channel’s spatial structure to outperform the conventional least-squares (LS) method, such an approach becomes computationally prohibitive when the number of antennas is large. This is especially true for uniform planar arrays (UPAs), where azimuth, elevation, and range parameters require a high-resolution three-dimensional grid search due to the narrow beamwidth and beamfocusing capabilities of ELAAs.

To address these challenges, we consider an ELAA with a modular UPA structure and propose a family of computationally efficient estimators that bypass the infeasibility of brute-force search while still leveraging the channel's spatial structure. Specifically, we develop constant-modulus LS (CM-LS) and reduced-subspace (RS) variants, and introduce a novel two-dimensional discrete Fourier transform (2D-DFT) masking technique that improves estimation performance and significantly reduces fronthaul signaling between the BBUs and the central processing unit (CPU). Numerical results confirm the robustness and effectiveness of the proposed methods, particularly in dense deployments and in the presence of LNA non-linearities.

        \vspace{-2mm}

\section{System Model}
        \vspace{-2mm}

\begin{figure}[t]  
    \centering
        \includegraphics[width=0.5\textwidth, trim=0cm 0cm 0cm 0cm, clip]{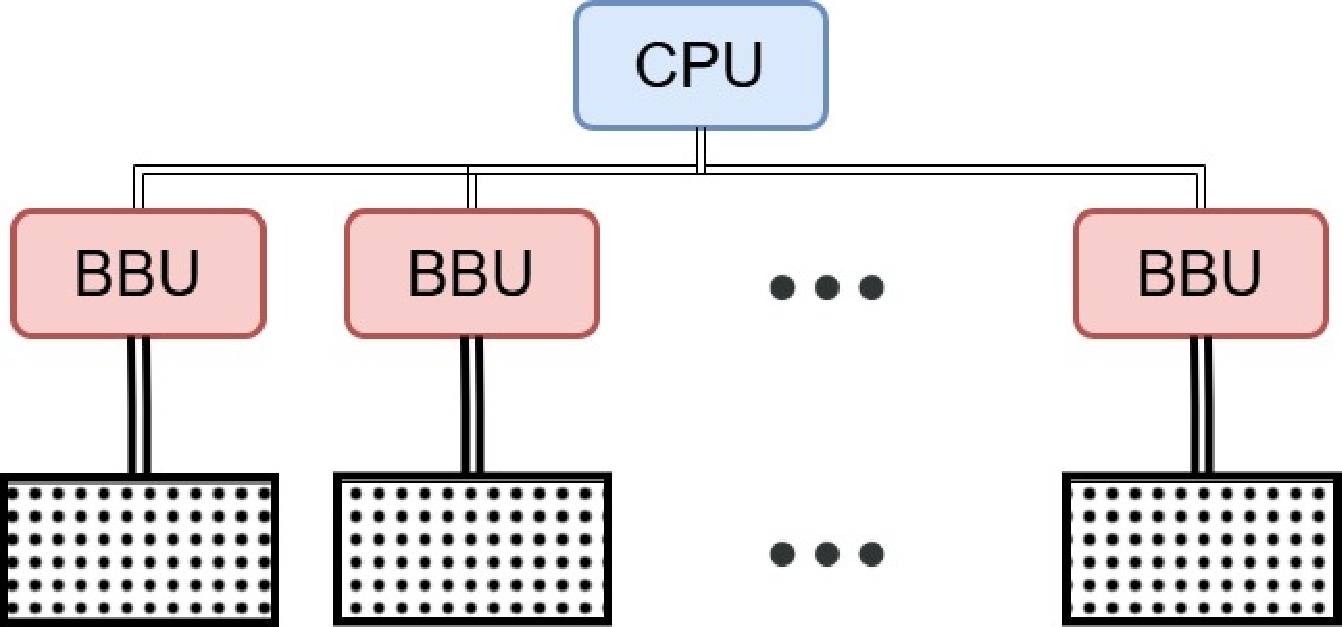} 
        \caption{Illustration of the modular ELAA system, where each subarray is configured as a UPA. }
        \label{fig:system}
        \vspace{-5mm}
 \end{figure}

We consider a modular ELAA architecture comprising $L$ subarrays as shown in Fig.~\ref{fig:system}. Each subarray consists of $N$ antennas arranged in a UPA and is connected to a dedicated BBU for local signal processing. The BBUs are coordinated by a CPU, which manages joint processing across the entire ELAA. In this work, we focus on the channel estimation for a single UE that has a single antenna. Channel estimation for multiple UEs can be performed by allocating separate pilot sequences. The UE transmits a pilot sequence, and the ELAA estimates the channel based on the received signal.

In this paper, we focus on the hardware distortions introduced by the LNAs connected to each antenna within a subarray. During uplink transmission, the received analog signals are first amplified by the LNAs, which introduces hardware-induced distortions due to the non-linear characteristics of the amplifiers. To isolate the impact of hardware impairments at the radio units, we assume ideal, error-free, and latency-free connections between the BBUs and the CPU.

We consider a UE located in the radiative near-field region of the ELAA and assume a LOS propagation environment. Each subarray is implemented as a UPA comprising $N_{\rm H}$ antennas along the horizontal axis and $N_{\rm V}$ antennas along the vertical axis, resulting in a total of $N = N_{\rm H} N_{\rm V}$ antennas per subarray. The inter-element spacing in both directions is denoted by $\Delta$. Antennas within each subarray are indexed in a row-by-row (raster-scan) fashion using an index $n \in \{1, \ldots, N\}$. Following the coordinate system defined in~\cite[Fig.~1]{emil_rayleigh_fading_ris}, the position of the $n$th antenna in subarray $l$, relative to the global origin, is given by the vector
$
\vect{u}_{l,n} = \begin{bmatrix} 0 \\ i_l(n)\Delta \\ j_l(n)\Delta \end{bmatrix}
$,
where $i_l(n)$ and $j_l(n)$ denote the horizontal and vertical indices of antenna $n$ in subarray $l$, respectively.

\begin{figure}[t]  
    \centering
        \includegraphics[width=0.5\textwidth, trim=1.5cm 0.2cm 1cm 1cm, clip]{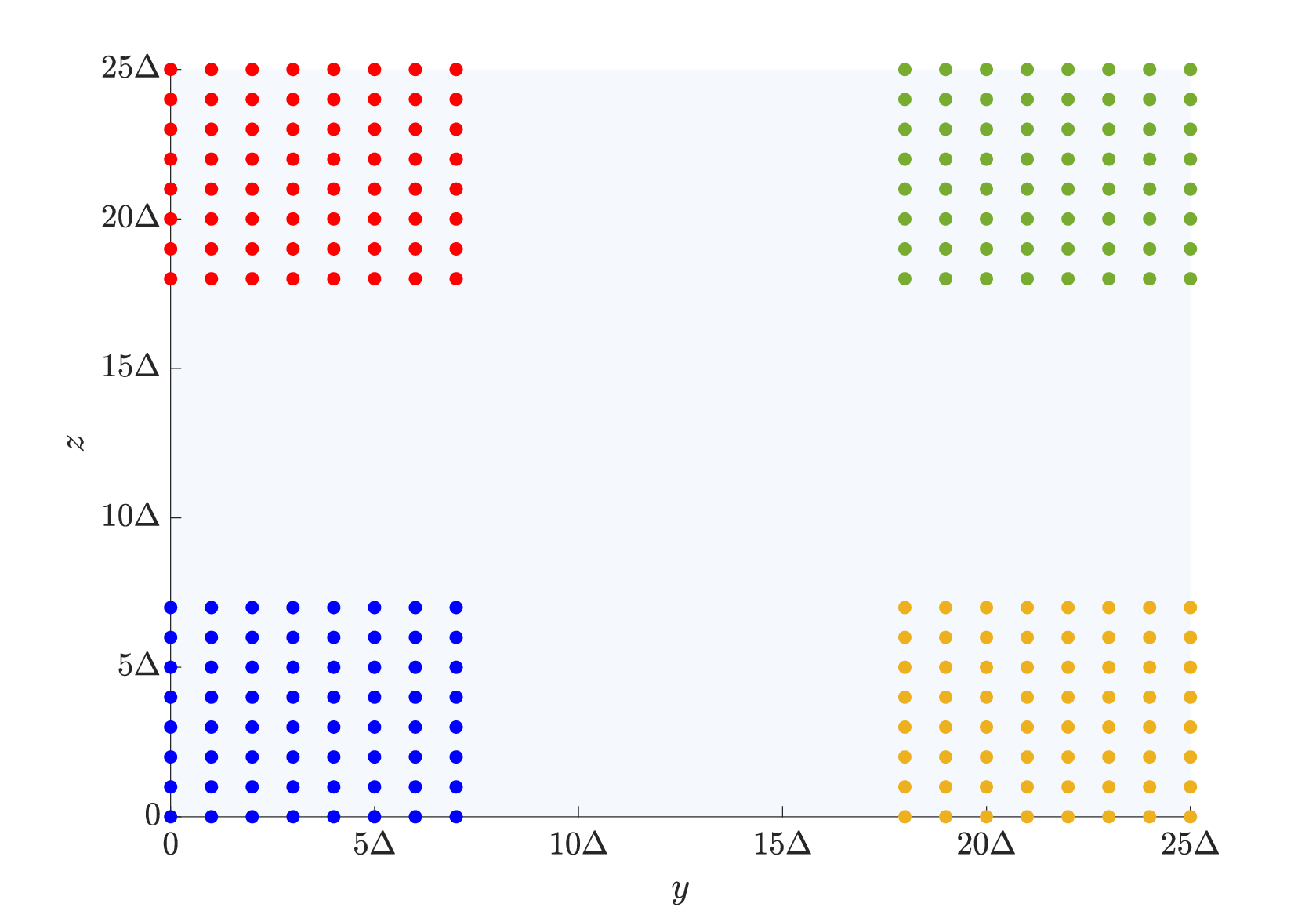} 
        \caption{Illustration of a modular ELAA system with $L=4$ subarrays, each subarray having $N=8\times 8=64$ antennas with horizontal and vertical $\overline{\Delta}=10\Delta$ separation. }
        \label{fig:modular}
                \vspace{-5mm}
 \end{figure}

We assume that the subarrays are deployed along the horizontal and vertical directions with a separation of $\overline{\Delta}$ between the edges of them. An example configuration with $L = 4$ subarrays, $N_{\rm H} = N_{\rm V} = 8$ antennas per dimension, and $\overline{\Delta} = 10\Delta$ is illustrated in Fig.~\ref{fig:modular}. 

The horizontal and vertical indices of the antennas in the bottom-left subarray (i.e., subarray~1) are given by
\begin{align}
  i_1(n) &= \mathrm{mod}(n - 1, N_{\rm H}), \\
  j_1(n) &= \left\lfloor \frac{n - 1}{N_{\rm H}} \right\rfloor,
\end{align}
where $\mathrm{mod}(\cdot,\cdot)$ denotes the modulo operation and $\lfloor \cdot \rfloor$ is the floor (truncation) operator.

On the other hand, the indices for subarray~4, located at the top-right corner of the ELAA, are given by
\begin{align}
  i_4(n) &= 18 + \mathrm{mod}(n - 1, N_{\rm H}), \\
  j_4(n) &= 18 + \left\lfloor \frac{n - 1}{N_{\rm H}} \right\rfloor.
\end{align}

The LOS channel of an arbitrary UE located in the radiative near-field region of the array is denoted by $\vect{h} \in \mathbb{C}^{LN}$. A wave arriving from the near-field has a spherical wavefront, implying that the channel can be modeled as in~\cite{cui2022channel}:
\begin{equation}
\vect{h} = \sqrt{\beta}
\begin{bmatrix}
\vect{b}_1\left(\varphi, \theta, r\right) \\
\vect{b}_2\left(\varphi, \theta, r\right) \\
\vdots \\
\vect{b}_L\left(\varphi, \theta, r\right)
\end{bmatrix},
\end{equation}
where $r$ denotes the distance between the origin (the reference corner of the ELAA) and the UE antenna, and $\beta > 0$ is the distance- and angle-dependent channel gain that captures both path loss and antenna directivity. The assumption of a common path loss coefficient across all antennas becomes tight for distances greater than twice the aperture of the array~\cite{bjornson2020power}.\footnote{The aperture of the ELAA refers to the length of its largest dimension. In the example shown in Fig.~\ref{fig:modular}, it is calculated as $D = 26\sqrt{2}\Delta$.}

The near-field array response vector $\vect{b}_l\left(\varphi, \theta, r\right)$ of subarray~$l$ depends on the azimuth and elevation angles $(\varphi, \theta)$, which are measured from the global origin. It characterizes the spherical wavefront as observed by subarray~$l$ and is expressed as
\begin{equation} \label{eq:near-field-array-response-vector}
\vect{b}_l\left(\varphi, \theta, r\right) = 
\left[
e^{-\imagunit\frac{2\pi}{\lambda}(r_{l,1}-r)}, \ldots, 
e^{-\imagunit\frac{2\pi}{\lambda}(r_{l,N}-r)}
\right]^{\Ttran},
\end{equation}
where $r_{l,n}$ is the distance between the $n$th antenna of subarray~$l$, located at $\vect{u}_{l,n} = [ 0, \, i_l(n)\Delta, \, j_l(n)\Delta ]^{\Ttran}$, and the UE. It is given by
\begin{align} \label{eq:distance-difference}
r_{l,n} &= \left(
\left[r\cos(\theta)\cos(\varphi)\right]^2 +
\left[r\cos(\theta)\sin(\varphi) - i_l(n)\Delta\right]^2 \right. \nonumber\\
&\quad \left. + \left[r\sin(\theta) - j_l(n)\Delta\right]^2
\right)^{\frac{1}{2}} \nonumber \\
&= r\Bigg( 1 
- \frac{2\Delta}{r}\big[i_l(n)\cos(\theta)\sin(\varphi) + j_l(n)\sin(\theta)\big] \nonumber \\
&\quad+ \frac{\Delta^2}{r^2}\left[i_l^2(n) + j_l^2(n)\right]
\Bigg)^{\frac{1}{2}}.
\end{align}

We assume that waves impinge on the array only from the front-facing hemisphere, i.e., $\varphi \in \left[-\frac{\pi}{2}, \frac{\pi}{2}\right]$.

In the following section, we introduce a signal model that incorporates impairments due to LNA non-linearities.

\subsection{Pilot Signal Distorted by LNA Nonlinearities}

In the uplink, the UE transmits a known pilot signal, which we assume to be the scalar value one without loss of generality. Denoting the uplink transmit power of the UE by $p$, the baseband-equivalent, distortion-free received signal at the input of the LNAs at the $N$ antennas of subarray~$l$ is given by
\begin{align}
\vect{y}_{l} = \sqrt{p \beta} \vect{b}_{l}(\varphi, \theta, r) + \vect{w}_l \in \mathbb{C}^{N}, \label{eq:received-data}
\end{align}
where $\vect{w}_l \in \mathbb{C}^{N}$ denotes the additive noise vector at subarray~$l$ during the pilot transmission phase. The entries of $\vect{w}_l$ are assumed to be independent and identically distributed as $\CN(0, \sigma^2)$. Moreover, the noise vectors $\vect{w}_l$ are mutually independent across different subarrays $l$.

Each antenna's RF signal is passed through the LNA dedicated to that antenna branch. For simplicity, we assume that all LNAs follow the same distortion model. The proposed techniques can easily be extended to the more generic case of different distortion models. Third-order polynomials with odd-order terms are widely used to model the compression effects in power amplifiers, as the third-order intercept point—related to the cubic term—is a common performance metric for RF amplifier distortion~\cite{Ronnow2019}. The quasi-memoryless model is valid when the bandwidth of the transmitted signal is sufficiently narrow compared to the total bandwidth of the amplifier~\cite{Jacobsson2018, Schenk2008a}.

 Using a third-order quasi-memoryless model, the distorted signal at the output of the $n$-th LNA of subarray $l$ at pilot transmission is given by
\begin{align}\label{eq:y-check}
\check{y}_{l,n} = a_{1} y_{l,n} + \frac{a_2}{\mathbb{E}\left\{ |y_{l,n}|^2 \right\}} \left| y_{l,n} \right|^2 y_{l,n},
\end{align}
where $y_{l,n} \in \mathbb{C}$ and $\check{y}_{l,n}\in \mathbb{C}$ denote the $n$-th entry of the received signal vector $\vect{y}_l$ and the LNA output vector $\check{\vect{y}}_{l}$, respectively. We assume that long-term automatic gain control (AGC) is applied at the receivers, following~\cite{demir2019channel}. The complex hardware coefficients $a_{1} \in \mathbb{C}$ and $a_{2} \in \mathbb{C}$ are specific to the characteristics of the employed LNA.

The expectation term $\mathbb{E}\left\{ |y_{l,n}|^2 \right\}$ in~\eqref{eq:y-check} can be computed as $
    \mathbb{E}\left\{ |y_{l,n}|^2 \right\} =  p \beta + \sigma^2$. Our goal is to estimate the normalized channel vector $\overline{\vect{h}} = \vect{h} / \sqrt{\beta}$ from the received pilot signal, assuming that the channel gain $\beta$ has already been estimated. Since the ELAA comprises a large number of antennas, $\beta$ can be accurately estimated using a sample variance-based approach. In the ideal case, a maximum likelihood channel estimator for jointly estimating the parameters $\varphi$, $\theta$, and $r$ can be designed \cite{abedin2010maximum}. However, such an estimator involves a substantial grid search complexity. In large-scale arrays, accurate estimation of angular and distance parameters becomes essential due to the narrow array beampattern, which results in a highly sensitive response. Achieving this level of precision typically requires a high-resolution three-dimensional grid search, which is computationally prohibitive.

The simplest alternative is the conventional LS estimator. However, it is often overly pessimistic, as it does not leverage the inherent structure of the channel or the geometry of the UPA. To address this limitation, we design a variant of the reduced-subspace least-squares (RS-LS) channel estimation method~\cite{demir2022channel} that explicitly accounts for the constant-modulus property of the entries in the array response vectors.

\section{Constant-Modulus  RS-LS Channel Estimation}
We first express the third-order distorted term as
\begin{align}
    |y_{l,n}|^2y_{l,n}= p\beta (\sqrt{p\beta}b_{l,n}(\varphi,\theta,r)+w_{l,n})+\check{w}_{l,n}
\end{align}
where $b_{l,n}(\varphi,\theta,r)$ and $w_{l,n}$ are the $n$th entries of $\vect{b}_{l,n}(\varphi,\theta,r)$ and $\vect{w}_l$. The additional distortion noise $\check{w}_{l,n}$ is obtained as
\begin{align}
    \check{w}_{l,n}& = \left(|w_{l,n}|^2+2\Re\left(\sqrt{p\beta}b_{l,n}(\varphi,\theta,r)w_{l,n}^*\right)\right) \nonumber\\
    &\quad \cdot\left(\sqrt{p\beta}b_{l,n}(\varphi,\theta,r)+w_{l,n}\right). 
\end{align}
Then, the impaired signal can be expressed as
\begin{align}
   \check{y}_{l,n} &= a_1\sqrt{p\beta}b_{l,n}(\varphi,\theta,r)+a_1w_{l,n} \nonumber\\
   &\quad+ \frac{a_2p\beta}{p\beta+\sigma^2}(\sqrt{p\beta}b_{l,n}(\varphi,\theta,r)+w_{l,n})+\frac{a_2}{p\beta+\sigma^2}\check{w}_{l,n} \nonumber\\
   &=\underbrace{\sqrt{p\beta}\left(a_1+a_2\frac{p\beta}{p\beta+\sigma^2}\right)}_{\triangleq \alpha\in \mathbb{C}}b_{l,n}(\varphi,\theta,r) \nonumber\\
  &\quad +\underbrace{\left(a_1+a_2\frac{p\beta}{p\beta+\sigma^2}\right)w_{l,n}+\frac{a_2}{p\beta+\sigma^2}\check{w}_{l,n}}_{\triangleq \dot{w}_{l,n}}. \label{eq:distorted}
\end{align}
Since the distortion model for each LNA is the same, we obtain the following model under hardware impairments:
\begin{align}
    \check{\vect{y}}_{l} = \alpha \vect{b}_{l}(\varphi, \theta, r) + \dot{\vect{w}}_l \in \mathbb{C}^{N},
\end{align}
where $\dot{\vect{w}}_l=[\dot{w}_{l,1} \ \cdots \ \dot{w}_{l,N}]^{\Ttran}$.

To implement the conventional LS estimator, the BBU of each subarray can independently estimate the channel, and the CPU can then combine all the subarray estimates. In this case, the LS estimate of the normalized channel $\overline{\vect{h}} = \vect{h}/\sqrt{\beta}$ is 
\begin{align}
\widehat{\overline{\vect{h}}}^{\mathrm{LS}} = \frac{1}{\alpha} \begin{bmatrix} \check{\vect{y}}_{1} \\ \vdots \\ \check{\vect{y}}_{L} \end{bmatrix}. \label{eq:LS}
\end{align}

Exploiting the constant-modulus property of the normalized channel vector, a better estimator can be constructed by retaining only the phase of each component. We refer to this estimator as the \emph{constant-modulus LS (CM-LS)} estimator:
\begin{align}
    \widehat{\overline{\vect{h}}}^{\mathrm{CM\text{-}LS}} = \exp\left(\imagunit \angle \left(\frac{1}{\alpha} \begin{bmatrix} \check{\vect{y}}_{1} \\ \vdots \\ \check{\vect{y}}_{L} \end{bmatrix} \right) \right), \label{eq:CM-LS}
\end{align}
where both $\exp(\cdot)$ and $\angle(\cdot)$ are applied element-wise. This estimator can be interpreted as the solution to a constrained LS problem, where the objective is to minimize the squared error subject to unit-modulus constraints on each element of the channel vector.

In~\cite{demir2022channel}, it is shown that when the eigenspace of the spatial correlation matrix corresponding to isotropic scattering (i.e., uniformly covering all angular directions) is considered, it spans all plausible channel directions. The orthogonal complement of this space—the null space of the isotropic eigenspace—contains only noise. This null space is non-empty for UPAs. This means that even if the geometry captures all physically plausible channel directions, some subspace components will never carry signal energy. While the results in~\cite{demir2022channel} assume ideal hardware, the same reasoning holds under third-order nonlinearities when the distorted signal is modeled as the array response vector scaled by $\alpha$ plus distortion noise, as in~\eqref{eq:distorted}.

Thus, noise in these null dimensions can be eliminated by projecting the received signal onto the valid channel subspace. Notably, this subspace is determined solely by the UPA geometry and spans all plausible LOS directions. Let $\vect{U} \in \mathbb{C}^{LN \times s}$ denote a tall semi-unitary matrix whose columns span the eigenspace of the isotropic spatial correlation matrix corresponding to nonzero eigenvalues, where $s \leq LN$ is the dimension of the reduced subspace. The \emph{reduced-subspace LS (RS-LS)} estimate is then given by
\begin{align}
    \widehat{\overline{\vect{h}}}^{\mathrm{RS\text{-}LS}} = \vect{U} \vect{U}^{\Htran}  \frac{1}{\alpha} \begin{bmatrix} \check{\vect{y}}_{1} \\ \vdots \\ \check{\vect{y}}_{L} \end{bmatrix},
\end{align}
which corresponds to projecting the noisy received signal onto the subspace where any valid channel vector can lie.

Taking this concept one step further, we propose the \emph{constant-modulus RS-LS (CM-RS-LS)} estimator, which additionally exploits the constant-modulus property:
\begin{align}
    \widehat{\overline{\vect{h}}}^{\mathrm{CM\text{-}RS\text{-}LS}} = \exp\left(\imagunit \angle \left( \vect{U} \vect{U}^{\Htran}  \frac{1}{\alpha} \begin{bmatrix} \check{\vect{y}}_{1} \\ \vdots \\ \check{\vect{y}}_{L} \end{bmatrix} \right) \right). \label{eq:CM-RS-LS}
\end{align}

To implement the CM-RS-LS estimator, each BBU receives the local signal and computes $\check{\vect{y}}_{l} / \alpha$ (or divides by subarray- and antenna-specific scalars if the distortion parameters vary across the array). The resulting normalized signals are then forwarded to the CPU, which performs the RS projection followed by the CM adaptation.

\section{Constant-Modulus DFT-Masked  Channel Estimator with Reduced Subspace Projection}
\label{sec:DFT}

To enhance noise rejection, can one design an improved estimator that leverages the unique characteristics of the near-field LOS channel, while still avoiding the computational burden of exhaustive grid search? Furthermore, is it possible to simultaneously reduce the signaling load from the BBUs to the CPU? To this end, we aim to develop a more efficient estimator that not only improves estimation accuracy but also minimizes the communication overhead to the CPU.

As demonstrated in~\cite{tang2024joint}, for uniform linear arrays (ULAs) under near-field LOS conditions, the energy of the DFT of the channel vector is concentrated around a dominant peak and a few bins representing the spatial bandwidth of the channel. A similar effect can be observed in UPAs by reshaping the channel vector into a two-dimensional (2D) matrix and computing its 2D-DFT. The resulting transform reveals that most of the channel energy is localized in a compact region of the 2D spatial frequency domain.

\begin{figure}[t]
    \centering

    \begin{subfigure}[b]{0.5\textwidth}
        \centering
        \includegraphics[width=\textwidth, trim=3cm 1cm 3cm 1cm, clip]{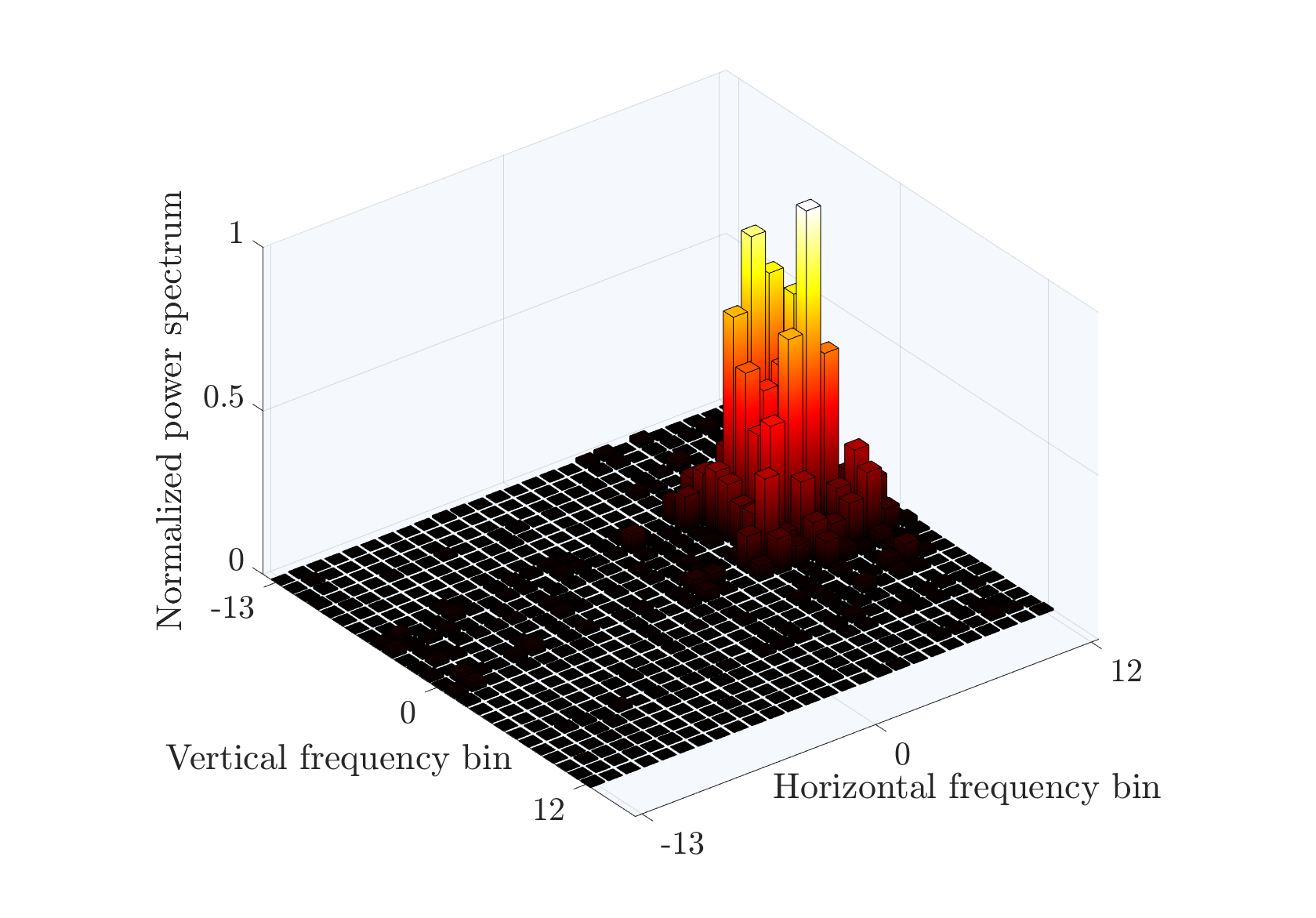}
        \caption{$\varphi=\pi/4$ and $\theta=0$.}
        \label{fig:bins1}
    \end{subfigure}

    \vspace{0.3cm} 

    \begin{subfigure}[b]{0.5\textwidth}
        \centering
        \includegraphics[width=\textwidth, trim=3cm 1cm 3cm 1cm, clip]{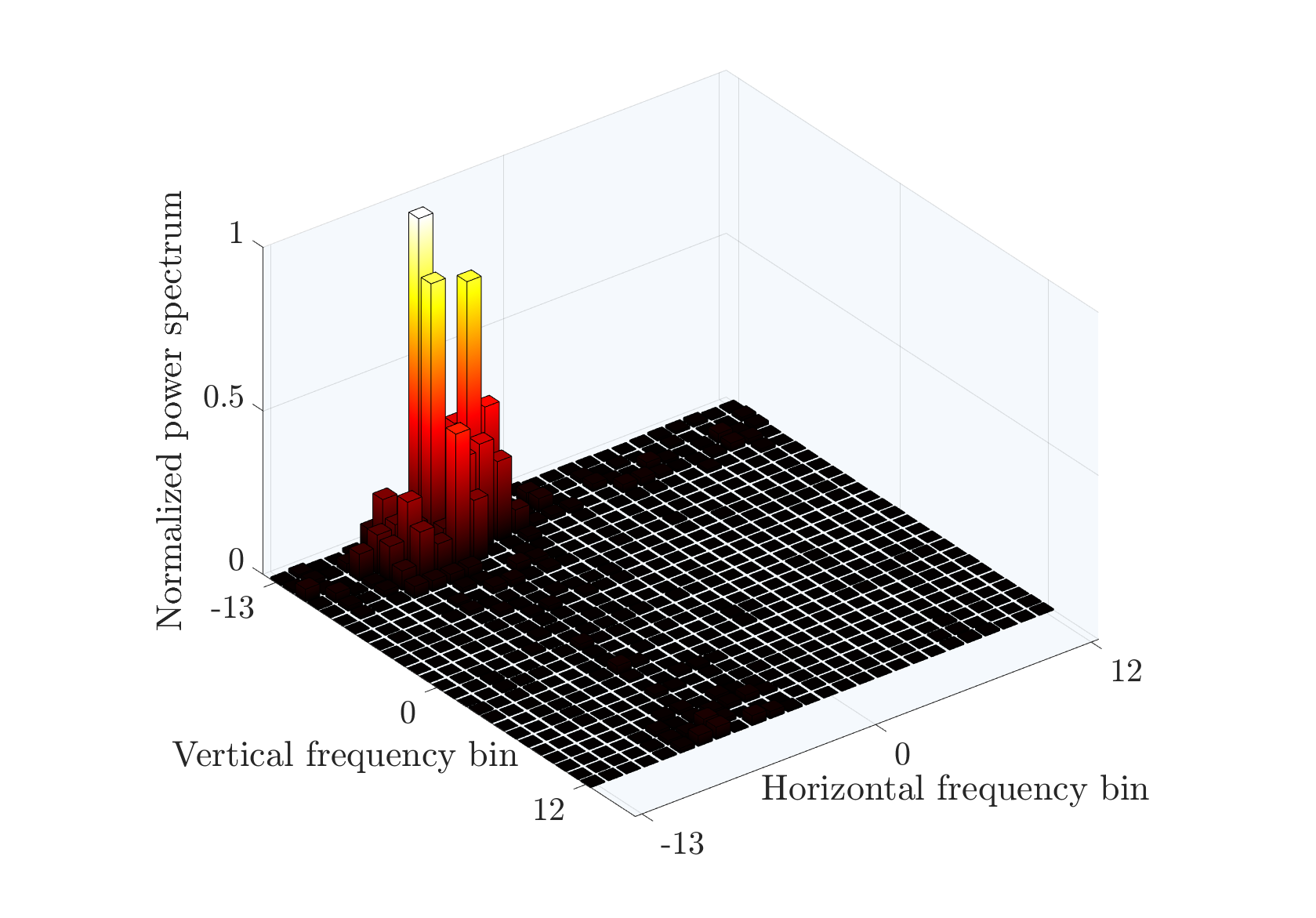}
        \caption{$\varphi=-\pi/4$ and $\theta=-\pi/3$.}
        \label{fig:bins2}
    \end{subfigure}

    \caption{Normalized 2D-DFT power spectrum of different near-field LOS channels for $r = 2D$, where the ELAA configuration corresponds to Fig.~\ref{fig:modular} and the array aperture is given by $D = 26\sqrt{2}\Delta$.
}
    \label{fig:bins}
            \vspace{-5mm}
\end{figure}

To illustrate this effect, we consider two different near-field LOS channels in Fig.~\ref{fig:bins}, based on the modular ELAA configuration shown in Fig.~\ref{fig:modular}. The aperture length is $D = 26\sqrt{2}\Delta$, where $\Delta = \lambda/2$ and the wavelength is $\lambda = 0.02\,\text{m}$, corresponding to a carrier frequency of $f_c = 15\,\text{GHz}$. The distance from the UE to the center of the ELAA is set to $r = 2D$. The received signal matrix is reshaped into a $26 \times 26$ grid, where the size follows from $2N_{\rm H} + 10 = 26$ and $2N_{\rm V} + 10 = 26$, with the value $10$ representing the inter-subarray separation given by $\overline{\Delta}/\Delta = 10$. After applying the 2D-DFT to the noisy and impaired signal and centralizing the zero-frequency component using \texttt{fftshift}, the spatial frequency spectrum is obtained. The SNR, which is $p\beta/\sigma^2$ is set to $10$\,dB. The third-order non-linearity parameters are $a_1=1.065$ and $a_2=-0.028$ \cite{Jacobsson2018}.

As outlined in~\cite{bjornson2024introduction}, the horizontal 2D-DFT bins correspond to the normalized horizontal spatial frequencies $\sin(\varphi)\cos(\theta)$, while the vertical bins correspond to the normalized vertical spatial frequency $\sin(\theta)$. The frequency bins range from $-13$ to $12$, which maps to the normalized spatial frequency interval $[-1, 1)$. Due to both the near-field energy spread effect and the finite resolution of DFT bins, energy is distributed around the dominant peak.

In Fig.~\ref{fig:bins}(a), the azimuth and elevation angles are set to $\varphi = \pi/4$ and $\theta = 0$, respectively. As a result, the energy is concentrated around a positive horizontal frequency bin (since $\sin(\varphi)\cos(\theta) > 0$) and zero vertical frequency bin ($\sin(\theta) = 0$). In contrast, Fig.~\ref{fig:bins}(b) corresponds to the case with $\varphi = -\pi/4$ and $\theta = -\pi/3$, where the energy is centered around negative horizontal and vertical frequency bins.

In the proposed technique, the BBU of each subarray first reshapes the LS estimator $\check{\vect{y}}_{1}/\alpha$ into a $N_{\rm H} \times N_{\rm V}$ matrix, corresponding to the 2D structure of the subarray. It then computes the 2D-DFT of this reshaped array. By identifying the most dominant components that account for $\delta$\% of the total 2D-DFT energy, only these significant frequency components are transmitted to the CPU. This not only reduces the fronthaul signaling load but also suppresses noise by discarding weak spatial components that are likely dominated by noise. The CPU reconstructs each subarray signal by applying the inverse 2D-DFT and concatenates all reconstructed signals into a received vector that matches the full ELAA dimension. Finally, the CM-RS-LS estimator, as defined in~\eqref{eq:CM-RS-LS}, is applied to this received vector to estimate the normalized channel.

In the next section, we quantitatively evaluate the performance of the various channel estimation techniques covered in this paper.

\section{Numerical Results}

In this section, we evaluate the performance of the channel estimation techniques in terms of the normalized mean squared error (NMSE). Six different channel estimation schemes are considered:  
\begin{itemize}
    \item[i)] \emph{LS} estimator from~\eqref{eq:LS};  
    \item[ii)] \emph{CM-LS} estimator from~\eqref{eq:CM-LS};  
    \item[iii)] \emph{CM-RS-LS} estimator from~\eqref{eq:CM-RS-LS};  
    \item[iv)] \emph{DFT-LS} estimator, which applies 2D-DFT masking as described in Section~\ref{sec:DFT} followed by the LS estimator in~\eqref{eq:LS};  
    \item[v)] \emph{DFT-CM-LS} estimator, which applies 2D-DFT masking followed by the CM-LS estimator in~\eqref{eq:CM-LS};  
    \item[vi)] \emph{DFT-CM-RS-LS} estimator, which applies 2D-DFT masking followed by the CM-RS-LS estimator in~\eqref{eq:CM-RS-LS}.
\end{itemize}

We consider $L=4$ subarrays deployed as illustrated in Fig.~\ref{fig:modular}, with vertical and horizontal spacing of $\overline{\Delta} = 10\Delta$. The wavelength is set to $\lambda = 0.02\,\text{m}$, corresponding to a carrier frequency of $f_c = 15\,\text{GHz}$. Each point in the performance curves represents the average over 1000 independent trials, where in each trial, the UE's distance from the ELAA center is randomly selected in the interval $[2D,\ 2D^2/\lambda]$ (i.e., up to the Fraunhofer distance), and the azimuth and elevation angles are drawn uniformly from $[-\pi/2,\ \pi/2]$. The third-order non-linearity parameters are set to $a_1 = 1.065$ and $a_2 = -0.028$, following~\cite{Jacobsson2018}. In the DFT masking process, an $\textrm{SNR}/(\textrm{SNR}+1)$ portion of the total 2D-DFT energy is preserved, where $\textrm{SNR}=p\beta/\sigma^2$.

\begin{figure}[t]  
    \centering
        \includegraphics[width=0.49\textwidth, trim=1cm 0.2cm 1cm 1cm, clip]{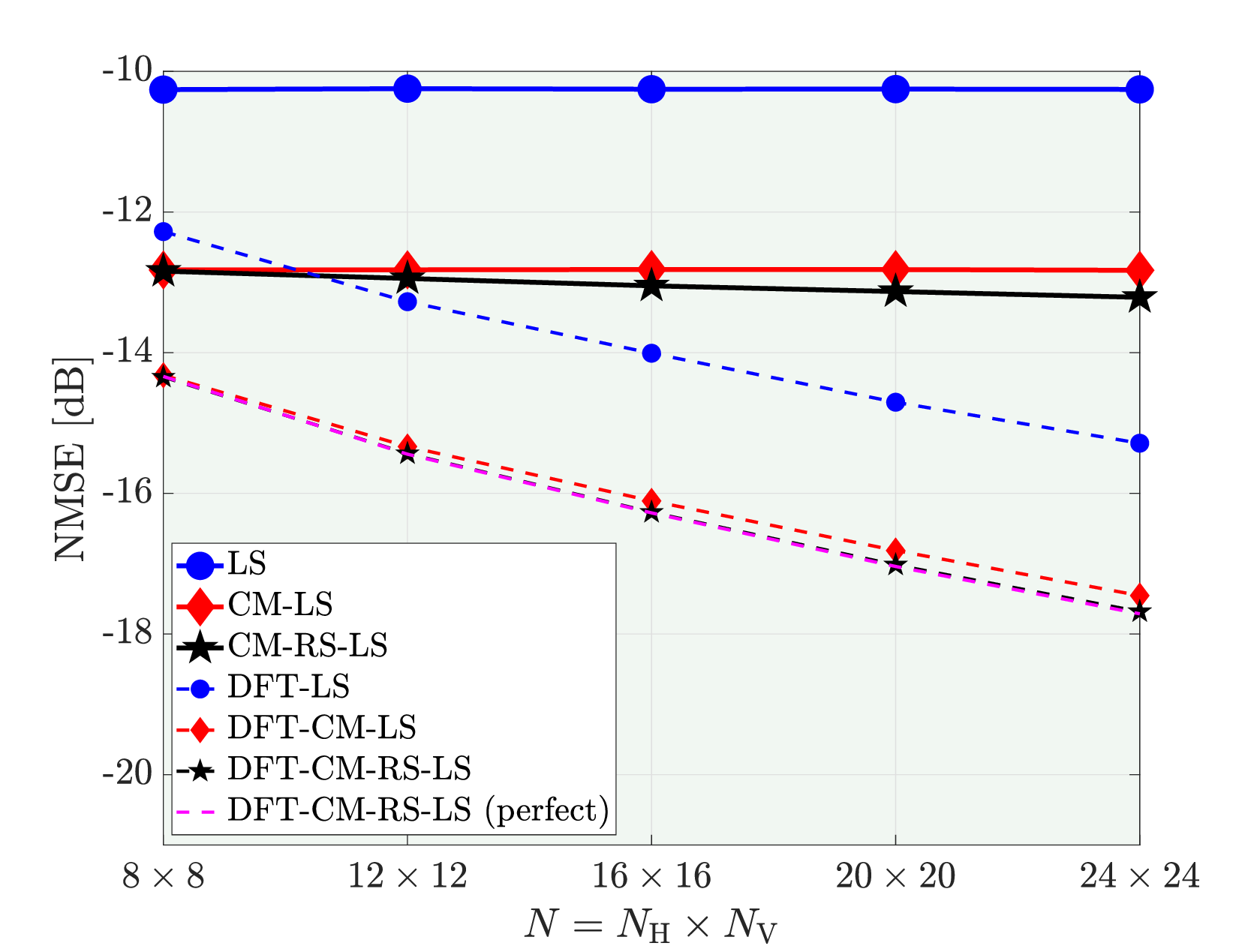} 
        \caption{The NMSE in terms of subarray size for $\Delta=\lambda/2$ and $\textrm{SNR}=10$\,dB. }
        \label{fig:4}
                \vspace{-5mm}
 \end{figure}

In Fig.~\ref{fig:4}, we fix the SNR at $10$\,dB and vary the subarray size along the horizontal axis. The inter-antenna spacing within each subarray is set to $\Delta = \lambda/2$. As observed from the figure, the performance of the LS and CM-LS estimators remains relatively stable with respect to the subarray size, with the CM-LS estimator consistently outperforming the conventional LS estimator. Furthermore, the CM-RS-LS estimator, which incorporates the RS projection, achieves superior performance, and its advantage over CM-LS increases as the subarray size $N$ grows. 

On the other hand, the application of DFT masking enhances the performance of all estimators. In particular, the performance of the DFT masking-based techniques improves with increasing subarray size. This enhancement is attributed to the ability of DFT masking to suppress noise in unused spatial directions by retaining only the significant 2D-DFT components. This behavior is further clarified in Table~\ref{table}, which presents the ratio of non-zero 2D-DFT components to the total number of subarray antennas. As the subarray size increases, the proportion of retained components decreases, effectively filtering out noise from irrelevant directions.

Moreover, with $\Delta = \lambda/2$, we observe that although RS projection provides a performance gain over its CM-LS counterpart, the improvement remains relatively modest. The performance of the DFT-CM-RS-LS estimator with perfect hardware is also shown as a reference and the impact of hardware impairments on the channel estimation performance is negligible.

\begin{figure}[t]  
    \centering
        \includegraphics[width=0.49\textwidth, trim=1cm 0.2cm 1cm 1cm, clip]{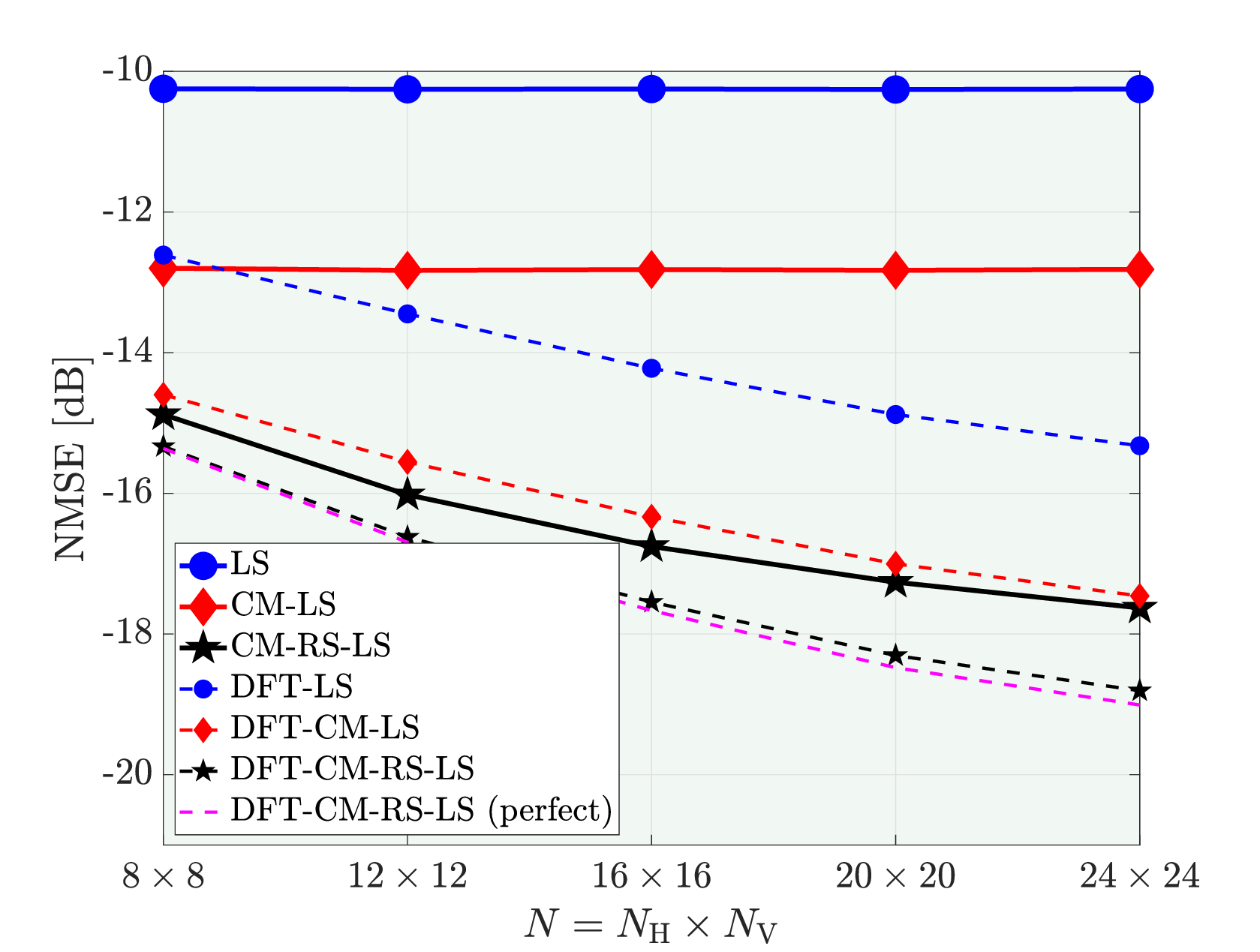} 
        \caption{The NMSE in terms of subarray size for $\Delta=\lambda/4$ and $\textrm{SNR}=10$\,dB. }
        \label{fig:5}
                \vspace{-5mm}
 \end{figure}

To better illustrate the advantage of RS projection, we reduce the inter-antenna spacing to $\Delta = \lambda/4$ and repeat the same experiment, as shown in Fig.~\ref{fig:5}. In this case, both the DFT-CM-RS-LS and CM-RS-LS estimators significantly outperform their respective CM-LS counterparts. This improvement is attributed to the oversampling effect introduced by the denser array, which results in a much smaller subset of effective channel dimensions relative to the total number of spatial samples, compared to the $\Delta = \lambda/2$ case~\cite{emil_rayleigh_fading_ris,demir2022channel}. Although the gap between perfect and imperfect hardware cases for DFT-CM-RS-LS estimator increases compared to the previous figure, the adverse impact of hardware impairments is still negligible with the proposed estimator.

\begin{table}[h!]
\caption{Proportion of non-zero 2D-DFT components relative to the number of antennas per subarray.}
\centering
\begin{tabular}{|c|c|c|c|c|c|c|}
\hline
\multicolumn{7}{|c|}{$N_{\rm H} \times N_{\rm V}$} \\
\hline
 &  & 8$\times$8 & 12$\times$12 & 16$\times$16 & 20$\times$20 & 24$\times$24 \\
\hline
\multirow{2}{*}{$\Delta$} & $\lambda/2$ & 0.1814
   &  0.1356
      &  0.1064
     & 0.0873
    &  0.0708 \\
\cline{2-7}
                          & $\lambda/4$ & 0.1765  & 0.1306  & 0.1049 & 0.0849  &  0.0694  \\
\hline
\end{tabular} \label{table}
        \vspace{-3mm}
\end{table}

In the implementation of the LS, CM-LS, and CM-RS-LS methods without DFT masking, the number of complex scalars that must be transmitted from the BBU to the CPU is $N$ for each channel estimation task. In contrast, for the DFT masking-based techniques, as illustrated in Table~\ref{table}, only a fraction of the total subarray size is transmitted, significantly reducing the fronthaul signaling load. This reduction becomes more pronounced as the arrays become denser and the number of antennas increases.

\section{Conclusion}
This paper investigated the problem of near-field LOS channel estimation for ELAAs with a modular UPA structure, under hardware impairments induced by third-order non-linearities in LNAs. We proposed a set of computationally efficient estimators that exploit both the constant-modulus structure of near-field LOS channels and the spatial characteristics imposed by the array geometry. To further reduce complexity and fronthaul signaling, we introduced a novel 2D-DFT masking technique that compresses the channel representation by retaining only the dominant frequency components. Numerical results demonstrated that the proposed estimators—particularly the CM-RS-LS estimator combined with DFT masking—achieve superior estimation accuracy, especially in dense array deployments. Our results highlight that modular ELAA architectures, coupled with structure-aware estimation techniques, are effective for practical near-field communications with scalable and low-cost hardware.

Moreover, when the LNA distortion model is accurately known and compensated at the subarray level, the detrimental effects of hardware impairments on channel estimation performance become negligible.

\bibliographystyle{IEEEtran}
\bibliography{IEEEabrv,refs}

\end{document}